\documentclass[11pt]{article}
\usepackage[margin=1in]{geometry}
\usepackage{booktabs}
\usepackage{float}
\usepackage{graphicx}
\usepackage{amsmath}
\usepackage{setspace}
\usepackage[hidelinks]{hyperref}
\usepackage{caption}
\usepackage[round]{natbib}
\usepackage{authblk}

\title{Good Question! \\[4pt]
{\normalfont\Large The Effect of Positive Feedback on Contributions to Online Public Goods}}

\author[1,2,3]{Johannes Wachs}
\author[4]{Leonore R\"oseler}
\author[5]{Tobias Gesche}
\author[5]{Elliott Ash}
\author[4,1,$\ast$]{Anik\'o Hann\'ak}

\affil[1]{Center for Collective Learning, Corvinus Institute of Advanced Studies\newline Corvinus University of Budapest, Hungary}
\affil[2]{Institute of Economics, ELTE Centre for Economic and Regional Studies, Hungary}
\affil[3]{Complexity Science Hub, Austria}
\affil[4]{Department of Informatics, University of Zurich, Switzerland}
\affil[5]{Center for Law \& Economics, ETH Zurich, Switzerland}
\affil[$\ast$]{Corresponding author: \texttt{hannak@ifi.uzh.ch}}

\date{}

\begin{document}
\maketitle

\begin{abstract}
\singlespacing \noindent Online platforms where volunteers answer each other's questions are important sources of knowledge, yet participation is declining. We ran a pre-registered experiment on Stack Overflow, one of the largest Q\&A communities for software development (N = 22,856), randomly assigning newly posted questions to receive an anonymous upvote. Within four weeks, treated users were 6.3\% more likely to ask another question and 12.9\% more likely to answer someone else's question. A second upvote produced no additional effect. The effect on answering was larger, more persistent, and still significant at twelve weeks. Next, we examine how much of these effects are due to algorithmic amplification, since upvotes also raise a question's rank and visibility. Algorithmic amplification is not important for the effect on asking additional questions, but it matters a lot for the effect on answering other questions. The increase in visibility increases the probability that another user provides an answer, and that experience appears to shift the poster toward broader community participation.
\end{abstract}

\newpage

\onehalfspacing

\section{Introduction}

Online knowledge platforms depend on voluntary contributions. Over fifteen years, Stack Overflow contributors have assembled one of the largest knowledge bases in software engineering, one consulted daily by millions of professionals and learners. This content now reaches far beyond the platform itself: the Stack Exchange network, of which Stack Overflow is the largest site, contributes roughly 3.4 times as much text by weight to The Pile, a widely used pre-training corpus for large language models, as Wikipedia does \citep{gao2020pile}. Yet contributions have been declining for years. The trend predates generative AI but has accelerated since the release of ChatGPT in late 2022 \citep{delriochanona2024,burtch2024ai}, as AI-assisted coding tools have spread rapidly through the software industry \citep{daniotti2026}. Related declines have been documented on Wikipedia \citep{halfaker2013} and other platforms where user-generated content is exposed to AI training \citep{peukert2024}. Understanding what sustains voluntary participation is a practical question for anyone who relies on these platforms, whether through a browser or through a language model.

A longstanding question in the study of public goods is what sustains voluntary contribution when downstream benefits are diffuse and contributors receive little direct return. One candidate is social feedback: a small signal that an effort was valued. Experimental evidence broadly supports this idea. Symbolic awards on Wikipedia, costly peer recognition on Reddit, and initial success signals on crowdfunding platforms have all been shown to increase contribution \citep{restivo2012,vanderijt2014,gallus2017,burtch2022}. Even small signals of social approval, it seems, can shift behavior. What remains unclear is whether anonymous, costless feedback on a single contribution produces comparable effects, and if so, through what channel.

Obtaining clean causal evidence is difficult, in part because the mechanism through which feedback operates is underspecified. On platforms with ranking algorithms, social feedback does not just send a psychological signal to the contributor; it also changes what happens next. An upvote on Stack Overflow alters the question's ranking, increases its visibility, and raises the probability that another user provides an answer. Any experiment that randomizes feedback on such a platform therefore conflates a direct social signal with algorithmically mediated social interaction. Prior recognition experiments have used badges and peer-to-peer awards on Wikipedia \citep{restivo2012,gallus2017} that do not feed into content-ranking algorithms, sidestepping this confound entirely. On real platforms, however, social feedback and algorithmic amplification are bundled together, and we treat this coupling as a feature to be modeled rather than noise to be assumed away.

We ran a pre-registered randomized controlled trial on Stack Overflow in which 22,856 users were randomly assigned to receive zero, one, or two anonymous upvotes on a recently posted question. The upvotes were indistinguishable from organic community feedback. Stack Overflow's scale, anonymity of voting, and one-to-one mapping between questions and users make it well-suited for this design. We tracked subsequent behavior over 4, 8, and 12 weeks, measuring outcomes separately for asking another question and for answering someone else's question.

The treatment increased both forms of participation. Treated users were 6.3\% more likely to ask another question within four weeks ($p < 0.05$) and 12.9\% more likely to answer one ($p < 0.01$); a second upvote added nothing beyond the first. Because votes affect ranking, the treatment also raised the question's visibility and its probability of receiving an answer from another user. We use several complementary approaches to bound how much of the treatment effect flows through this algorithmically mediated channel, in which the upvote raises visibility, attracts an answer, and the answer changes behavior, versus the direct channel of the upvote signal itself. The two channels contribute in strikingly different proportions to the two outcomes: the direct channel accounts for the majority of the effect on asking, while the mediated channel accounts for a substantial share of the effect on answering. The effect on asking attenuates by twelve weeks; the effect on answering persists, consistent with the idea that receiving substantive help from another user shifts behavior more durably than the upvote signal alone.

This study makes three contributions. First, it provides experimental evidence that anonymous peer feedback on a single contribution increases the recipient's subsequent participation in an online public good, including answering other users' questions. Second, it demonstrates a decomposition of the treatment effect into a direct social signal and an algorithmically mediated pathway, showing that their relative contributions differ sharply across outcomes. Third, it documents a spillover from receiving feedback on a question to answering other users' questions, linking low-cost social feedback to the kind of prosocial behavior that sustains knowledge platforms.

\section{Setting and Experimental Design}

This study was approved by the Human Subjects Committee of the University of Zurich (OEC IRB \#2021-103) and pre-registered on AsPredicted.org (ID: \href{https://aspredicted.org/f72g-k8jn.pdf}{96592}).

\subsection{Stack Overflow}

Stack Overflow is a question-and-answer (Q\&A) platform for programming, where users post questions, provide answers, and vote on each other's contributions. The resulting net vote count is prominently displayed next to each post. Votes serve two functions: they rank content within the platform's search and display algorithms, and they aggregate into user-level reputation scores. Reputation unlocks privileges (commenting, editing, voting) and acts as a public signal of expertise. Each upvote on a question awards the poster 10 reputation points; for the median user in our sample, whose baseline reputation is 15, a single upvote represents a 67\% increase.

This design creates a tight coupling between social feedback and content visibility. An upvote simultaneously sends a signal to the poster and changes how the platform distributes attention to the question. This coupling is central to our study: it means that randomizing an upvote intervenes on both channels at once.

\subsection{The Experiment}

The experiment ran for 121 days, from June 20 through October 19, 2022. Four times daily (at 00:00, 06:00, 12:00, and 18:00 UTC), an automated script scanned Stack Overflow's questions feed, which averaged roughly 3,600 new questions per day during this period. From each scan, 80 questions were randomly selected to receive upvotes and a further 200 to serve as controls, yielding a daily target of 280 observations. Of the 80 treated questions, half received one upvote and half received two. The first upvote arrived within six hours of the question's posting. The second, where applicable, was staggered by two hours to match the typical arrival rate of organic feedback.

The target sample size was 33,600 question-user pairs (4,800 single-treated, 4,800 double-treated, 24,000 control). The unequal allocation across conditions reflects power calculations, the goal of treating fewer than 2\% of daily new questions to minimize platform interference, and capacity constraints in the sampling procedure. Any user who had previously been sampled was skipped, so all observations map one-to-one to a distinct user. Users cannot see who upvotes their posts, ruling out effects driven by the upvoter's identity. Users were unaware of their participation; consent is governed by Stack Overflow's terms of service, which authorize public display and use of all user actions. Several months after the experiment ended, we deleted all accounts used for upvoting, removing all experimental votes and leaving no long-run distortion.

The last question in our sample was collected on October 19, 2022, and the primary outcome for this final observation was recorded on November 15, 2022. ChatGPT was released on November 30, 2022. Our data therefore predate the sharp decline in platform engagement that followed the release of generative AI tools.

\subsection{Sample and Attrition}

The final sample contains 22,856 question-user pairs. Attrition from the initial target of 33,600 is driven primarily by question deletion between sampling and follow-up (approximately 21\% of cases), with smaller contributions from changes in user or page identifiers (2.8\%) and time-outs (2.7\%). Attrition is higher in the control arm (32.8\%) than the double-treated arm (28.5\%), producing statistically significant differential attrition ($\chi^2 = 42.1$, $p < 0.001$; Table~\ref{tab:s5}). This has a mechanical consequence: unanswered questions are more likely to be deleted, so the surviving control sample contains a larger share of questions that already had an answer at baseline (20.7\% in control vs.\ 14.7\% in single-treated and 13.8\% in double-treated). Pre-treatment user-level covariates are otherwise balanced across arms (Table~\ref{tab:s6}). \citet{lee2009} bounds confirm that the treatment effects are robust to worst-case selective attrition (Table~\ref{tab:s7}). All primary outcomes survive Bonferroni, Benjamini--Hochberg, and Holm multiplicity corrections at $\alpha = 0.05$ (Table~\ref{tab:s9}).

\subsection{Outcome Variables}

The primary, pre-registered outcome measures whether each user was active within four weeks of posting the focal question. We measure two forms of activity separately: whether the user asked another question and whether the user answered another user's question. We also measure any activity (asking or answering). We extended data collection to eight and twelve weeks to examine persistence (Table~\ref{tab:s1}).

All log-transformed control variables use $\log(x + 1)$: specifically, prior questions posted, prior answers posted, and question views at baseline. The ratio increase of views is defined as $\text{Views}_{t_4} / \text{Views}_{t_0}$, where $t_0$ is the time of treatment assignment and $t_4$ is the four-week follow-up. All questions have at least one view at baseline, so the denominator is always positive. ``Receives an answer'' is an indicator for whether the question gained at least one new answer between baseline and the endpoint.

\section{Related Work}

\subsection{Motivations for contributing to online public goods}

Why do people contribute to platforms where the benefits flow mostly to others? On Q\&A sites, the most direct motive is getting one's own question answered, but only a minority of askers return to participate beyond that initial exchange \citep{bachschi2020}. For those who stay, the reasons are varied: building expertise through practice \citep{vasilescu2013,vasilescu2014}, signaling skills to potential employers \citep{xu2020,forderer2024}, and accumulating reputation through gamified systems that make these signals visible \citep{cavusoglu2015,anderson2013,moldon2021}. Alongside these self-interested motives runs a prosocial thread. Some contributors are drawn by the satisfaction of helping others \citep{vadlamani2020}, some by the desire for recognition and esteem \citep{benabou2006}, and some by what \citet{andreoni1990} calls ``warm-glow,'' the private benefit of giving itself. The balance between these motives is delicate: monetary rewards tend to crowd out the intrinsic ones \citep{gneezy2000,kheramnuai2018}, while symbolic recognition can reinforce them \citep{kosfeld2011}.

Two theoretical traditions bear directly on our findings. The first is generalized reciprocity: the idea that receiving help increases a person's willingness to help unrelated third parties, even when there is no possibility of direct repayment \citep{nowak2005,stanca2009}. The second is social identity theory, which holds that individuals who come to identify with a group internalize its norms, including norms of mutual help \citep{tajfel1979,akerlof2000}. \citet{gallus2017} interprets the effect of symbolic awards on Wikipedia editors in these terms. Both traditions predict that positive feedback could shift users from narrow, self-interested participation toward broader community contribution. That prediction maps closely onto the pattern we observe.

\subsection{Experimental evidence on recognition and contributions}

Field experiments have consistently shown that recognition increases contributions to online public goods, but the form of recognition matters. Awards and tokens that carry visible status implications produce the largest effects. On Wikipedia, symbolic awards increase editor retention by roughly 20\% \citep{gallus2017}, and peer-to-peer recognition tokens raise productivity by 60\%, an effect that persists over 90 days \citep{restivo2012}. On Reddit, Gold awards (which are visible, costly, and peer-initiated) increase posting volume, though they also steer recipients toward content similar to what was rewarded \citep{burtch2022}. Across crowdfunding, ratings, and petition platforms, \citet{vanderijt2014} find that a small initial success triggers cascading advantages, though with diminishing marginal returns. The broader lesson from this literature is that social approval shifts behavior, and that the shift is larger when the signal is public and carries reputational weight.

What we know much less about is whether anonymous, costless feedback works through similar channels. The closest precedent is \citet{muchnik2013}, who randomize anonymous votes on a social news site but measure herding in \textit{others'} voting rather than the recipient's own behavior. Studies of newcomer feedback on Wikipedia and Slashdot find positive effects on retention \citep{lampe2005,farzan2012,zhu2013}, but those interventions are delivered by identifiable accounts and results vary with implementation details. 

Few experiments attempt to isolate what happens to the recipient of an anonymous, low-cost signal of approval, which are the most common forms of feedback users get on these communities. In a concurrent and independent experiment, \citet{jiang2025} randomizes anonymous upvotes on Stack Overflow \textit{answers} among already-active answerers and finds a 15\% increase in subsequent answering, with no effect on question-asking. Our experiment differs in three ways: we upvote questions rather than answers, we sample askers (including a substantial share of first-time posters), and our mechanism question is not social versus instrumental motivation but the direct signal versus algorithmic amplification.

\subsection{Algorithmic amplification}

On platforms with content-ranking algorithms, a vote does more than signal approval. It changes what other users see. The ``Music Lab'' experiments of \citet{salganik2006} demonstrated this forcefully: when popularity was visible, small initial differences in song quality were amplified into vast inequalities in downloads, far exceeding what quality alone would predict. \citet{muchnik2013} showed the same dynamic at smaller scale, finding that a single positive vote on a social news site inflated final ratings by 25\% through herding. \citet{vanderijt2014} documented similar success-breeds-success patterns across four different platforms. The implication for our setting is that randomizing an upvote on Stack Overflow does not simply send a signal to the poster; it also changes the question's trajectory through the platform's information architecture, potentially attracting answers and attention that would not otherwise have arrived. Prior recognition experiments on Wikipedia sidestepped this issue because badges and awards do not feed into content-ranking algorithms. Our experiment cannot sidestep it, and so we treat the entanglement between social feedback and algorithmic amplification as a central object of study.

\subsection{AI and the sustainability of knowledge platforms}

The rise of large language models has given new urgency to questions about platform sustainability. The scale of dependence is substantial: \citet{vincent2019ugc} show that user-generated content from Wikipedia, Stack Overflow, and similar platforms appears in the vast majority of search engine results, and \citet{vincent2021datalev} argue that this dependence gives contributors a form of ``data leverage'' that could, in principle, be exercised collectively. In practice, however, the leverage runs in the other direction. Stack Overflow activity declined sharply after the release of ChatGPT \citep{delriochanona2024}, though the pattern is not uniform across platforms: \citet{burtch2024ai} find that Reddit programming communities, which have stronger social ties, showed no comparable drop. On Unsplash, content creators reduced uploads and left the platform at higher rates after their photographs were included in an AI training dataset \citep{peukert2024}. \citet{taraborelli2015} anticipated this dynamic, describing a ``paradox of reuse'': the more useful platform content becomes to external systems, the less reason users have to visit the platform and engage with the community that produced it. Whether AI will always require fresh human-generated content is an open question, but declining participation threatens the value of these platforms to human users regardless of what language models need.

\section{Empirical Analysis}

\subsection{Main results}

Pre-treatment covariates are balanced across treatment arms (Table~\ref{tab:s6}). The differential attrition discussed in Section 2 creates a mechanical imbalance in one baseline variable (the share of questions already answered at baseline), but \citet{lee2009} bounds confirm that treatment effects are robust to worst-case selective attrition (Table~\ref{tab:s7}).

\begin{figure}
\centering
\includegraphics[width=0.85\textwidth]{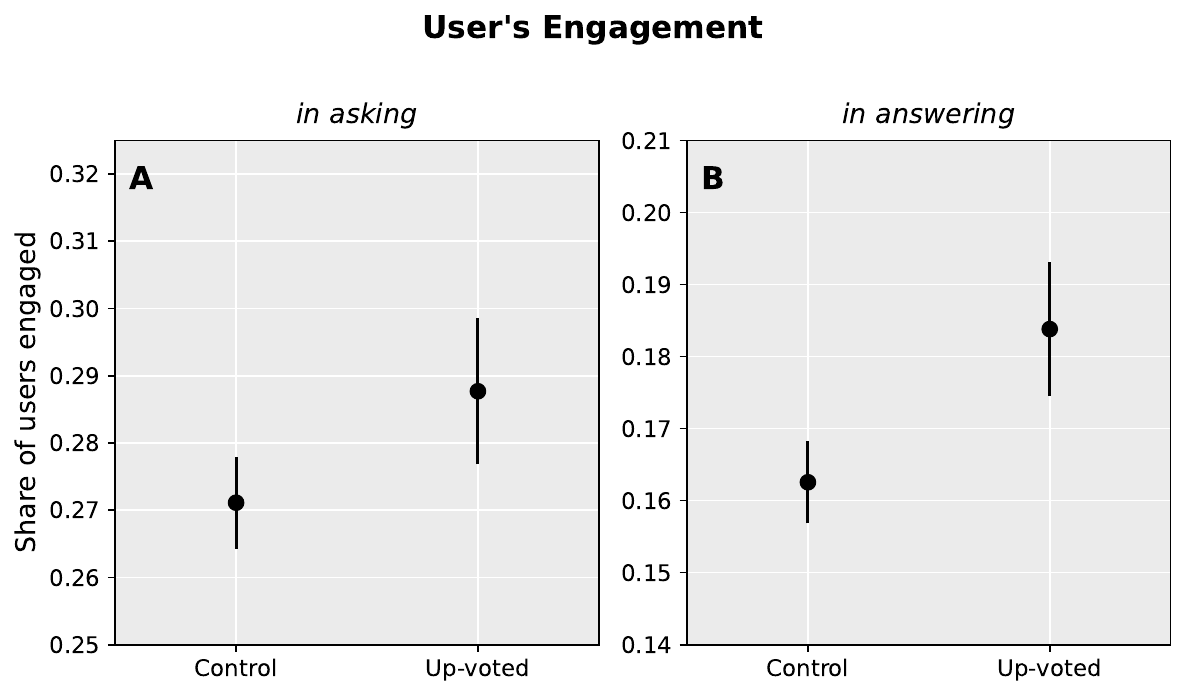}
\caption{Share of users engaged in asking (Panel A) and answering (Panel B) within four weeks, by treatment status (pooled). Points show group proportions with 95\% confidence intervals. The difference is significant for both outcomes ($p < 0.05$ for asking, $p < 0.01$ for answering).}
\label{fig:engagement}
\end{figure}

Our primary outcome is user activity four weeks after posting the focal question. Figure~\ref{fig:engagement} compares pooled treatment and control groups. Panel A reports the probability of posting a new question: 27.1\% of control users ask again, compared with 28.8\% of treated users, a 6.3\% increase. Panel B shows that upvotes also increase the probability of answering another user's question: 16.3\% of control users answer at least one question within four weeks, compared with 18.4\% of treated users, a 12.9\% increase.

To test the statistical significance of these differences and account for heterogeneity in pre-existing user attributes, we use a pre-registered regression model:
\[
\text{Engagement}_i = \beta_1 \cdot \text{Treated}_i + \beta_2 \cdot \text{DoubleTreated}_i + \gamma \mathbf{X}'_i + \epsilon_i,
\]
where $i$ indexes focal questions, each mapping one-to-one to a user. $\text{Engagement}_i$ is a binary outcome measuring whether the user made another contribution within four weeks. $\text{Treated}_i$ indicates whether the question received any treatment upvotes (single or double). $\text{DoubleTreated}_i$ equals 1 only when the question received two upvotes, capturing the additional effect of a second upvote relative to a single one. We control for the log number of questions and answers previously posted and the log number of views the question received prior to observation.

\begin{table}
\centering
\small
\resizebox{\textwidth}{!}{
\begin{tabular}{l ccc ccc ccc}
\toprule
 & \multicolumn{3}{c}{New Question} & \multicolumn{3}{c}{New Answer} & \multicolumn{3}{c}{New Post} \\
 & M1 & M2 & M3 & M4 & M5 & M6 & M7 & M8 & M9 \\
\midrule
Treated & 0.017$^{*}$ & 0.025$^{**}$ & 0.033$^{***}$ & 0.021$^{***}$ & 0.024$^{**}$ & 0.030$^{***}$ & 0.027$^{***}$ & 0.035$^{***}$ & 0.047$^{***}$ \\
 & (0.007) & (0.009) & (0.009) & (0.006) & (0.007) & (0.007) & (0.007) & (0.009) & (0.009) \\
Double Treated & & $-$0.016 & $-$0.014 & & $-$0.006 & $-$0.009 & & $-$0.016 & $-$0.018 \\
 & & (0.011) & (0.011) & & (0.009) & (0.009) & & (0.012) & (0.012) \\
Controls & & & Yes & & & Yes & & & Yes \\
\midrule
Control mean & 0.271 & 0.271 & 0.271 & 0.163 & 0.163 & 0.163 & 0.361 & 0.361 & 0.361 \\
$N$ & 22,856 & 22,856 & 22,856 & 22,856 & 22,856 & 22,856 & 22,856 & 22,856 & 22,856 \\
\bottomrule
\end{tabular}
}
\caption{Treatment effect on user behavior (4 weeks). The dependent variable is whether the user posted a new question (M1--M3), answer (M4--M6), or either (M7--M9) within four weeks. Linear probability model with HC1 robust standard errors in parentheses. Controls are log prior questions, log prior answers, and log baseline views. $^{*}p<0.05$, $^{**}p<0.01$, $^{***}p<0.001$. $N = 22{,}856$.}
\label{tab:main}
\end{table}

Table~\ref{tab:main} reports the results. The unconditional estimate (M1) shows a 1.7 percentage point increase in the probability of posting another question ($p < 0.05$), corresponding to a 6.3\% increase relative to the control mean. Separating single and double upvotes (M2) reveals that single-treated users are 2.5 percentage points more likely to ask again, a 9.2\% increase ($p < 0.01$), with no additional effect of a second upvote ($\beta_2 = -0.016$, $p > 0.1$). Adding controls does not meaningfully change these patterns (M3). The same structure holds for answering (M4 through M6): the pooled treatment effect is 2.1 percentage points, a 12.9\% increase ($p < 0.01$), again with no marginal effect of a second upvote. Combining both outcomes, the treatment increases any engagement by 2.7 percentage points, or 7.5\% ($p < 0.01$; M7). First-time posters, who make up 29.5\% of the sample, respond no differently from experienced users (Table~\ref{tab:s3}).

\begin{figure}
\centering
\includegraphics[width=\textwidth]{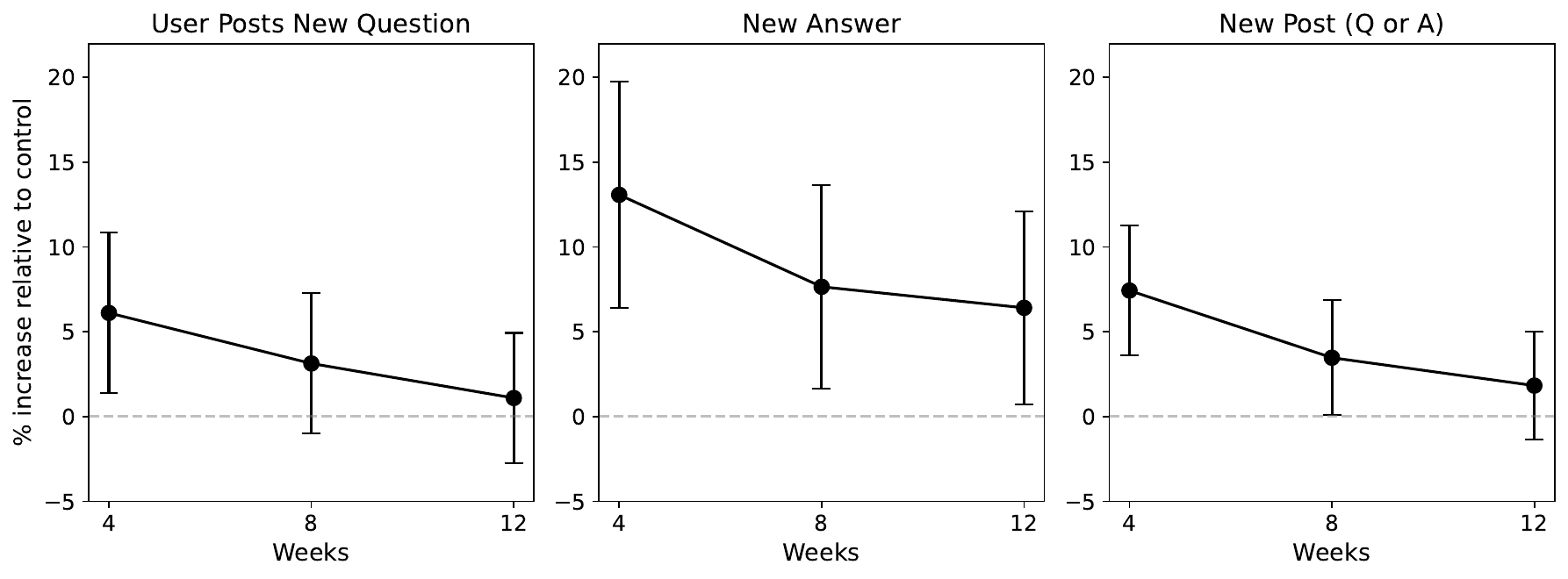}
\caption{Treatment effect over time. Points show the percentage increase in each outcome relative to the control mean at 4, 8, and 12 weeks, with 95\% confidence intervals. The asking effect attenuates to insignificance by 12 weeks. The answering effect remains significant throughout.}
\label{fig:decay}
\end{figure}

We repeat the analysis at eight and twelve weeks (Figure~\ref{fig:decay}). The effect on asking decays from 6.1\% at four weeks to an insignificant 1.1\% at twelve, while the effect on answering proves more durable: it declines from 12.9\% to 7.7\% at eight weeks and 6.4\% at twelve, remaining statistically significant throughout. The second upvote adds nothing at any horizon, though the study is underpowered to detect small differences between single and double treatment (achieved power 10--30\%; Table~\ref{tab:s8}), so this null should be read as an absence of evidence rather than evidence of absence.

\subsection{Mechanism analysis}

An upvote on Stack Overflow has two consequences: it signals appreciation to the poster, and it raises the question's ranking, increasing visibility and the probability that another user provides an answer. The treatment effect could flow through either channel, and we use several approaches to characterize how much flows through each.

Treatment increases answer receipt, and treated questions are more likely to receive an answer (Table~\ref{tab:questions}). Among the 18,564 users whose questions had not yet been answered at baseline, treatment increases the probability of receiving an answer by 7.9 percentage points ($p < 0.001$). This pathway operates through substantive human response, not mere exposure: among control users, receiving an answer strongly predicts future participation, while receiving more views without an answer does not (Table~\ref{tab:s13}). The treatment is also null where this chain is muted: among users whose questions already had an answer at baseline ($N = 4{,}292$), treatment has no detectable effect on any outcome (Table~\ref{tab:s10}). These are not simply underpowered nulls; equivalence tests reject effects larger than $\pm 5$ percentage points in this subgroup ($p < 0.02$ for all outcomes).

\begin{table}
\centering
\small
\begin{tabular}{l cc cc}
\toprule
 & \multicolumn{2}{c}{Receives an Answer} & \multicolumn{2}{c}{Ratio Increase of Views} \\
 & M1 & M2 & M3 & M4 \\
\midrule
Treated & 0.089$^{***}$ & 0.093$^{***}$ & 1.31$^{***}$ & 1.53$^{***}$ \\
 & (0.007) & (0.010) & (0.39) & (0.57) \\
Double Treated & & $-$0.009 & & $-$0.42 \\
 & & (0.013) & & (0.71) \\
\midrule
Control mean & 0.558 & 0.558 & 5.06 & 5.06 \\
$N$ & 22,856 & 22,856 & 22,856 & 22,856 \\
\bottomrule
\end{tabular}
\caption{Effects on questions (4 weeks). Treated questions are substantially more likely to receive an answer and receive more views. ``Receives an Answer'' is an indicator for gaining at least one new answer between baseline and four weeks. ``Ratio Increase of Views'' is the ratio of views at four weeks to views at baseline. HC1 robust standard errors in parentheses. $^{***}p<0.001$. $N = 22{,}856$.}
\label{tab:questions}
\end{table}

But this algorithmically mediated pathway --- in which the upvote raises visibility, attracts an answer, and the answer changes behavior --- cannot explain the full treatment effect, particularly for asking. We can test this by asking: if answer receipt were the \textit{only} channel, how large would its causal effect need to be? Dividing the treatment effect on behavior by the treatment effect on answer receipt gives an implied causal effect of receiving an answer. For asking, this implied effect is 5.0 times larger than the observed association between answer receipt and asking among untreated users, an association that is itself likely inflated by selection. In other words, for amplification alone to account for the asking result, receiving an answer would need to be implausibly potent. For answering, the implied effect is only 1.5 times the observational benchmark, consistent with the mediated pathway playing a larger role in that outcome (Table~\ref{tab:wald}).

\begin{table}
\centering
\begin{tabular}{l rrrrr}
\toprule
Outcome & ITT & Wald & Obs.\ Assoc. & Ratio \\
\midrule
New Question & 0.027$^{***}$ & 0.339 & 0.068 & 5.0$\times$ \\
New Answer & 0.023$^{***}$ & 0.290 & 0.187 & 1.5$\times$ \\
New Post & 0.039$^{***}$ & 0.491 & 0.199 & 2.5$\times$ \\
\midrule
\multicolumn{5}{l}{First stage: $\hat{\pi} = 0.079^{***}$ (SE $= 0.008$)} \\
\bottomrule
\end{tabular}
\caption{Can algorithmic amplification explain the treatment effect? Sample restricted to questions not yet answered at baseline ($N = 18{,}564$). ITT is the effect of treatment on subsequent asking or answering activity. Wald = ITT/$\hat{\pi}$: the implied effect of receiving an answer, if amplification were the sole channel. Obs.\ Association is the observed relationship between answer receipt and later activity among untreated users. Ratio = Wald/Obs.\ Association. A ratio well above 1 means that amplification alone requires implausibly large effects of answer receipt to explain the treatment effect.}
\label{tab:wald}
\end{table}

Both pieces of evidence point in the same direction: the direct channel dominates for asking, while the mediated channel is substantial for answering. Under additional assumptions, we can put rough numbers on the split. If the observational association upper-bounds the causal effect of answer receipt, the mediated channel can account for on the order of 20\% of the effect on asking and perhaps as much as two-thirds of the effect on answering (Table~\ref{tab:bounds}). A parametric mediation analysis \citep{imai2010} yields consistent proportions (Table~\ref{tab:s14}), and a sensitivity analysis confirms that the mediation result for answering is the more robust of the two: modest mediator-outcome confounding ($\rho = 0.08$) would reduce the asking mediation to zero, while a substantially larger correlation ($\rho = 0.24$) would be needed to eliminate the answering result. Additional supporting evidence from predicted-answerability terciles and the never-answered subgroup appears in Tables S15 and S11.

\begin{table}
\centering
\begin{tabular}{l rrrr}
\toprule
Outcome & ITT & Mediated share ($\leq$) & 95\% CI & Direct share ($\geq$) \\
\midrule
New Question & 0.027 & 20.0\% & [9.7\%, 30.3\%] & 80.0\% \\
New Answer & 0.023 & 64.6\% & [31.0\%, 98.1\%] & 35.4\% \\
New Post & 0.039 & 40.5\% & [24.9\%, 56.2\%] & 59.5\% \\
\bottomrule
\end{tabular}
\caption{Bounding the amplification share. Upper bound on the share of the ITT attributable to the algorithmically mediated channel, assuming the controlled observational association among untreated users equals the causal effect of answer receipt. All quantities are computed within the not-yet-answered subsample ($N = 18{,}564$). CIs are delta-method approximations. The observational association is likely an overestimate of the true causal effect, making these generous upper bounds on mediation.}
\label{tab:bounds}
\end{table}

\section{Discussion and Conclusion}

In this study, we find that a single anonymous upvote on a Stack Overflow question increases subsequent participation: treated users are 6.3\% more likely to ask another question and 12.9\% more likely to answer someone else's question within four weeks. These two effects differ not only in magnitude but in character. The effect on answering is larger, persists for at least twelve weeks, and is substantially mediated by the algorithmically amplified pathway through which the upvote raises a question's visibility and its probability of receiving an answer from another user. The effect on asking is smaller, fades by twelve weeks, and appears to be driven predominantly by the direct signal of the upvote itself. A second upvote adds nothing beyond the first.

The most striking result is not that an upvote leads to more question-asking but that it also leads to answering. This is hard to explain on self-interested grounds. Once a user's question has been answered, the problem that brought them to the platform is solved. Treatment does not create any instrumental reason to start answering the questions of others, yet that is what we observe.

What could account for this? One possibility is generalized reciprocity: receiving help from one person makes you more willing to help someone else, even someone unrelated to the original exchange \citep{nowak2005,stanca2009}. On this account, the answer activates a prosocial impulse that gets redirected toward the community at large. A second possibility is that the experience changes how users see themselves. Someone who receives a thoughtful answer may begin to feel like a member of a community rather than a consumer of a service, and that shift in identity brings with it the community's norms of mutual help \citep{akerlof2000,gallus2017}. A third possibility is simpler: returning to the site to read the answer creates a habit. The additional visits lower the cost of future participation, and the behavior persists after the original reason for visiting has passed \citep{charness2009}.

We cannot cleanly distinguish among these accounts, but the data do narrow the field. The 12-week persistence of the answering effect fits identity and habit formation better than a fleeting reciprocity impulse, which would be expected to fade quickly. The finding that high view counts without answer receipt show no association with later participation tells us that mere exposure to the platform is not enough; what matters is the substantive human response. And the absence of stronger effects among first-time posters makes it difficult to sustain a narrow onboarding story in which the upvote mainly reassures newcomers. The most defensible reading is that answer receipt is a consequential intermediate step, one whose behavioral consequences extend well beyond self-interested use of the platform.

The asking effect tells a different story. Here the evidence points toward a direct effect of the upvote itself, with the algorithmically mediated pathway playing a smaller role. What exactly the upvote conveys is less clear. It could be a signal that someone noticed the question, or the notification it generates could simply bring the user back to the site. It could also be the 10 reputation points, which for the median user represent a 67\% increase and which carry both gamification \citep{anderson2013} and signaling \citep{xu2020} value. That a second upvote adds nothing is consistent with a threshold account: one acknowledgment is enough, and a second crosses no additional boundary \citep{kosfeld2011,vanderijt2014}. It is also possible that the marginal visibility from a second vote is too small to matter in Stack Overflow's ranking algorithm.

The distinction between these two channels has practical consequences, and they pull in different directions. The direct signal tells us that simply making it easy for users to express appreciation has behavioral consequences on its own, independent of any algorithmic machinery. The ranking infrastructure matters for a different reason: it connects questioners with answerers and generates the substantive interactions that appear to sustain longer-term participation. A platform that invests in one channel while neglecting the other captures only part of the potential benefit. This distinction is especially relevant as AI interfaces begin to mediate access to platform content. When users consume answers through a language model rather than visiting the site, they generate fewer votes, fewer answers, and less of the social feedback that keeps contributors engaged \citep{taraborelli2015,delriochanona2024}. Finally, because our experiment predates ChatGPT, the upvotes were unambiguously read as appreciation from other humans, giving us a clean estimate of what small social signals can do before AI mediation reshaped what feedback on these platforms means. Whether feedback delivered through AI products can credibly substitute for that lost human signal, and counteract the post-ChatGPT decline in participation, is a concrete design priority for follow-up work.

These findings come with important caveats. The treatment bundles a social signal, algorithmic amplification, and a small reputation gain into a single intervention. Our analyses bound how much the mediated channel can explain, but they cannot disentangle the direct signal from the reputation gain or the notification, and the bounds rest on assumptions whose sensitivity we report. The general mechanism likely applies on other online platforms, but the effects and their relative sizes likely differ based on these platforms' implementation details, such as whether and how their reputation system is gamified. We observe only the extensive margin (whether a user contributes at all) over 4 to 12 weeks, and we cannot speak to contribution quality, longer-run adaptation, or the equilibrium responses that might follow if upvoting were systematically encouraged at scale.

Future experiments could pull apart what the upvote bundles together: sending a notification without changing the ranking, boosting visibility without sending a signal, or awarding reputation without a vote. Such designs would likely require collaboration with the platform, but they would tell us which component does the most work. More broadly, sustaining voluntary knowledge production requires understanding which feedback loops matter most, and how they can be preserved as the platforms that host them continue to change.

\bigskip
\noindent\textbf{Acknowledgments.} We thank L\'aszl\'o Czaller, Zoltan Elekes, S\'andor Juh\'asz, Gerg\H{o} T\'oth, and the members of the UZH Social Computing Group including Joachim Baumann, Azza Bouleimen, Corinna Hertweck, Stefania Ionescu, Nicol\`o Pagan, Zachary Roman, and Aleksandra Urman, as well as Stefan Menzel and Bernhard Sendhoff for helpful comments and suggestions. LR and AH gratefully acknowledge financial support from Honda Research Institute Europe (HRI-EU). JW acknowledges support from the Hungarian National Scientific Fund (OTKA FK 145960) and was partially funded by the European Union under Horizon EU project LearnData (101086712).

\bigskip
\noindent\textbf{Data, Materials, and Software Availability.} Anonymized data and a single consolidated analysis script that reproduces all main and supplementary tables and figures are available on Zenodo at \href{https://doi.org/10.5281/zenodo.19485377}{10.5281/zenodo.19485377}.

\clearpage

\appendix
\section*{Appendix}
\renewcommand{\thetable}{S\arabic{table}}
\renewcommand{\thefigure}{S\arabic{figure}}
\setcounter{table}{0}
\setcounter{figure}{0}

\subsection*{Sample and design}

Tables~\ref{tab:s1}--\ref{tab:s4} describe the experimental sample and report the main treatment effects at extended time horizons and for subgroups. Table~\ref{tab:s1} compares observed to expected group sizes. Table~\ref{tab:s2} extends the main results to 12 weeks. Table~\ref{tab:s3} tests for differential effects among first-time posters. Table~\ref{tab:s4} reports question-level outcomes (answer receipt, views, votes).

\begin{table}[H]
\centering
\begin{tabular}{lccc}
\toprule
Experiment Group & Observed & Expected & Contribution to $\chi^2$ \\
\midrule
Control & 16,137 & 16,326 & 2.18 \\
Single-Treated & 3,287 & 3,265 & 0.15 \\
Double-Treated & 3,432 & 3,265 & 8.53 \\
\bottomrule
\end{tabular}

\vspace{4pt}
\caption{Observed and Expected Group Sizes with Contributions to Chi-square \textit{Note:} $\chi^2 = 10.85$, $p$-value $< 0.01$.}
\label{tab:s1}
\end{table}

\begin{table}[H]
\centering
\resizebox{\textwidth}{!}{
\small
\begin{tabular}{l ccc ccc ccc}
\toprule
 & \multicolumn{3}{c}{New Question} & \multicolumn{3}{c}{New Answer} & \multicolumn{3}{c}{New Post} \\
 & M1 & M2 & M3 & M4 & M5 & M6 & M7 & M8 & M9 \\
\midrule
Treated (yes) & 0.004 & 0.012 & 0.015 & 0.014$^{*}$ & 0.020$^{*}$ & 0.027$^{***}$ & 0.009 & 0.018 & 0.023$^{*}$ \\
 & (0.007) & (0.010) & (0.009) & (0.006) & (0.009) & (0.008) & (0.008) & (0.010) & (0.010) \\
Double-treated (yes) & & $-$0.016 & $-$0.013 & & $-$0.011 & $-$0.013 & & $-$0.018 & $-$0.017 \\
 & & (0.012) & (0.012) & & (0.011) & (0.010) & & (0.013) & (0.012) \\
Prior Questions & & & 0.115$^{***}$ & & & $-$0.007$^{*}$ & & & 0.078$^{***}$ \\
 & & & (0.003) & & & (0.003) & & & (0.003) \\
Prior Answers & & & $-$0.012$^{***}$ & & & 0.033$^{***}$ & & & 0.009$^{***}$ \\
 & & & (0.001) & & & (0.001) & & & (0.001) \\
Question Votes at $t_0$ & & & 0.036$^{***}$ & & & 0.011 & & & 0.028$^{***}$ \\
 & & & (0.008) & & & (0.007) & & & (0.008) \\
\midrule
Mean Control Outcome & 0.378 & 0.378 & 0.378 & 0.222 & 0.222 & 0.222 & 0.470 & 0.470 & 0.470 \\
Observations & 20,118 & 20,118 & 20,118 & 20,118 & 20,118 & 20,118 & 20,118 & 20,118 & 20,118 \\
\bottomrule
\end{tabular}
}

\vspace{4pt}
\caption{Estimates of the Effect of Treatment and Double Treatment on User Posting Behavior after 12 Weeks. Specifically: the likelihood that a user will post a new question (M1--M3), answer (M4--M6), or either (M7--M9) within twelve weeks of observation. Estimates are derived using a linear probability model fit via OLS. We report heteroscedasticity-robust standard errors in parentheses. $^{*} = p < .05$, $^{**} = p < .01$, $^{***} = p < .001$.}
\label{tab:s2}
\end{table}

\begin{table}[H]
\centering
\begin{tabular}{lccc}
\toprule
 & New Question & New Answer & New Post \\
\midrule
Treated (yes) & 0.015 & 0.025$^{***}$ & 0.027$^{**}$ \\
 & (0.008) & (0.007) & (0.008) \\
First-timer (yes) & $-$0.152$^{***}$ & $-$0.120$^{***}$ & $-$0.204$^{***}$ \\
 & (0.007) & (0.005) & (0.008) \\
Treated (yes) $\times$ First-timer (yes) & 0.008 & $-$0.012 & $-$0.002 \\
 & (0.013) & (0.010) & (0.014) \\
Question Votes & 0.035$^{***}$ & 0.024$^{***}$ & 0.042$^{***}$ \\
 & (0.006) & (0.005) & (0.007) \\
Intercept & 0.313$^{***}$ & 0.195$^{***}$ & 0.417$^{***}$ \\
 & (0.004) & (0.004) & (0.005) \\
\midrule
Observations & 22,856 & 22,856 & 22,856 \\
$R^2$ & 0.024 & 0.023 & 0.039 \\
\bottomrule
\end{tabular}

\vspace{4pt}
\caption{Estimates of the Effect of Treatment and Double Treatment for First-timers. Outcomes are whether the users posted another question, answer, or either kind of post within 4 weeks of observation. We report robust standard errors. $^{*} = p < .05$, $^{**} = p < .01$, $^{***} = p < .001$.}
\label{tab:s3}
\end{table}

\begin{table}[H]
\centering
\small
\resizebox{\textwidth}{!}{
\begin{tabular}{l cc cc cc cc}
\toprule
 & \multicolumn{2}{c}{Has Answer} & \multicolumn{2}{c}{N. Answers} & \multicolumn{2}{c}{N. Views} & \multicolumn{2}{c}{N. Votes} \\
 & M1 & M2 & M3 & M4 & M5 & M6 & M7 & M8 \\
\midrule
Treated (yes) & 0.089$^{***}$ & 0.093$^{***}$ & 0.136$^{***}$ & 0.152$^{***}$ & 12.42$^{***}$ & 14.58$^{***}$ & 0.218$^{***}$ & 0.214$^{***}$ \\
 & (0.007) & (0.010) & (0.012) & (0.017) & (2.37) & (3.21) & (0.015) & (0.020) \\
Double-treated (yes) & & $-$0.009 & & $-$0.031 & & $-$4.20 & & 0.009 \\
 & & (0.013) & & (0.023) & & (4.92) & & (0.029) \\
\midrule
Mean Control Outcome & 0.558 & 0.558 & 0.947 & 0.947 & 75.36 & 75.36 & $-$0.03 & $-$0.03 \\
Observations & 22,856 & 22,856 & 22,856 & 22,856 & 22,856 & 22,856 & 22,856 & 22,856 \\
\bottomrule
\end{tabular}
}

\vspace{4pt}
\caption{Estimates of the Effect of Treatment and Double Treatment on Question-Level Outcomes within 4 weeks. The dependent variables are characteristics of the focal question measured four weeks after observation. We report heteroscedasticity-robust standard errors in parentheses. $^{*} = p < .05$, $^{**} = p < .01$, $^{***} = p < .001$.}
\label{tab:s4}
\end{table}

\newpage
\subsection*{Attrition, balance, and multiplicity}

Tables~\ref{tab:s5}--\ref{tab:s9} address threats from differential attrition, covariate imbalance, and multiple testing. Table~\ref{tab:s5} documents attrition rates by arm. Table~\ref{tab:s6} reports covariate balance. Table~\ref{tab:s7} applies Lee (2009) bounds to the primary outcomes. Table~\ref{tab:s8} assesses statistical power for the single-vs.-double treatment comparison. Table~\ref{tab:s9} applies multiplicity corrections.

\begin{table}[H]
\centering
\begin{tabular}{lrrrr}
\toprule
Stage & Control & Single & Double & Total \\
\midrule
Target (randomized) & 24,000 & 4,800 & 4,800 & 33,600 \\
Observed (analyzed) & 16,137 & 3,287 & 3,432 & 22,856 \\
Lost & 7,863 & 1,513 & 1,368 & 10,744 \\
Attrition rate & 32.8\% & 31.5\% & 28.5\% & 32.0\% \\
\bottomrule
\end{tabular}

\vspace{4pt}
\caption{CONSORT-Style Attrition Flow. The experiment targeted 33,600 question-user pairs. Attrition occurred primarily through question deletion between sampling and four-week follow-up. A chi-square test for differential attrition across arms is reported below. \textit{Note:} Chi-square test for differential attrition: $\chi^2 = 33.95$, $df = 2$, $p < 0.001$. OLS regression of retention on treatment indicators (with robust standard errors): Single coefficient $= 0.012$ ($p = 0.09$); Double coefficient $= 0.043$ ($p < 0.001$). Joint $F$-test: $F = 17.82$, $p < 0.001$. The double-treated arm retains 4.3 percentage points more observations than the control arm. Lee bounds (Table S7) address potential bias from this differential attrition.}
\label{tab:s5}
\end{table}

\begin{table}[H]
\centering
\resizebox{\textwidth}{!}{
\small
\begin{tabular}{l rrr rr rr}
\toprule
 & & & & \multicolumn{2}{c}{ANOVA} & \multicolumn{2}{c}{$p$ (vs.\ Control)} \\
\cmidrule(lr){5-6} \cmidrule(lr){7-8}
Variable & Control & Single & Double & $F$ & $p$ & Single & Double \\
\midrule
\textit{User-level covariates} & & & & & & & \\
Prior questions & 19.16 & 18.51 & 18.52 & 0.25 & 0.780 & 0.598 & 0.590 \\
Prior answers & 28.70 & 19.35 & 33.86 & 1.98 & 0.139 & 0.094 & 0.396 \\
Prior reputation & 1,322 & 1,007 & 1,490 & 1.38 & 0.251 & 0.175 & 0.487 \\
Account age (days) & 1,202 & 1,194 & 1,185 & 0.26 & 0.774 & 0.738 & 0.497 \\
First-time poster & 0.30 & 0.29 & 0.30 & 0.04 & 0.964 & 0.785 & 0.954 \\
\addlinespace
\textit{Question-level covariates (post-treatment)} & & & & & & & \\
Question views ($t_0$) & 10.59 & 8.88 & 8.85 & 232.6 & $<$0.001 & $<$0.001 & $<$0.001 \\
Question votes ($t_0$) & $-$0.03 & $-$0.04 & $-$0.04 & 2.16 & 0.116 & 0.260 & 0.062 \\
Question length (chars) & 527 & 544 & 540 & 1.75 & 0.174 & 0.112 & 0.204 \\
Question has code & 0.68 & 0.66 & 0.67 & 2.61 & 0.073 & 0.024 & 0.445 \\
Question n.\ comments & 0.64 & 0.46 & 0.47 & 41.43 & $<$0.001 & $<$0.001 & $<$0.001 \\
Question n.\ answers & 0.26 & 0.19 & 0.17 & 58.46 & $<$0.001 & $<$0.001 & $<$0.001 \\
\bottomrule
\end{tabular}
}

\vspace{4pt}
\caption{Baseline Balance across Treatment Arms. Means of pre-treatment covariates by experimental group, with one-way ANOVA $F$-statistics and pairwise $t$-tests versus control. User-level covariates (prior questions, prior answers, reputation, account age, first-time poster) are balanced across arms. Question-level covariates (views, comments, answers at $t_0$) show significant imbalance because they are measured after treatment and reflect the mechanical effect of upvotes on question visibility. \textit{Note:} Question-level imbalances in views, comments, and number of answers reflect post-treatment contamination: upvoted questions receive more visibility through Stack Overflow's ranking algorithms, which crowds out organic engagement. These variables are not included as controls in the main specifications for this reason.}
\label{tab:s6}
\end{table}

\begin{table}[H]
\centering
\begin{tabular}{l rrr rr}
\toprule
 & Lower & Naive & Upper & Trim & Trim \\
Outcome & bound & effect & bound & \% & group \\
\midrule
\textit{Panel A: Pooled treatment vs.\ control} & & & & & \\
Asked another question & $-$0.013 & 0.017 & 0.028 & 3.9\% & treated \\
Answered another question & $-$0.012 & 0.021 & 0.029 & 3.9\% & treated \\
Any new post & 0.002 & 0.027 & 0.043 & 3.9\% & treated \\
\addlinespace
\textit{Panel B: Single-treated vs.\ control} & & & & & \\
Asked another question & 0.012 & 0.025 & 0.030 & 1.8\% & treated \\
Answered another question & 0.009 & 0.024 & 0.028 & 1.8\% & treated \\
Any new post & 0.024 & 0.035 & 0.042 & 1.8\% & treated \\
\addlinespace
\textit{Panel C: Double-treated vs.\ control} & & & & & \\
Asked another question & $-$0.037 & 0.009 & 0.027 & 6.0\% & treated \\
Answered another question & $-$0.033 & 0.018 & 0.030 & 6.0\% & treated \\
Any new post & $-$0.021 & 0.019 & 0.043 & 6.0\% & treated \\
\bottomrule
\end{tabular}

\vspace{4pt}
\caption{Lee (2009) Bounds for Differential Attrition. Bounds are computed by trimming the excess observations from the group with higher retention. If both bounds share the same sign, the treatment effect is robust to worst-case selection. All outcomes measured at 4 weeks. \textit{Note:} Panel B: All three outcomes have positive lower bounds, confirming that the single-treatment effect is robust to worst-case attrition-driven selection. Panel A: The composite ``any new post'' outcome is robust; questions and answers individually have lower bounds near zero. Panel C: Double-treated bounds include zero for all outcomes, consistent with the higher differential attrition in this arm (6.0\% trimming) and the null incremental effect of the second upvote reported in the main text.}
\label{tab:s7}
\end{table}

\begin{table}[H]
\centering
\begin{tabular}{l rr rr rr}
\toprule
 & \multicolumn{2}{c}{MDE (pp)} & \multicolumn{2}{c}{Actual} & \multicolumn{2}{c}{Achieved} \\
\cmidrule(lr){2-3} \cmidrule(lr){4-5} \cmidrule(lr){6-7}
Outcome & 80\% & 90\% & Diff (pp) & Cohen's $d$ & Power \\
\midrule
Asked another question & 3.1 & 3.6 & 1.6 & 0.035 & 0.30 \\
Answered another question & 2.7 & 3.1 & 0.6 & 0.015 & 0.10 \\
Any new post & 3.3 & 3.9 & 1.6 & 0.034 & 0.28 \\
\bottomrule
\end{tabular}

\vspace{4pt}
\caption{Power Analysis for the Single-vs.-Double Treatment Contrast. The minimum detectable effect (MDE) is computed at 80\% and 90\% power ($\alpha = 0.05$, two-sided) given the observed sample sizes in each arm. Actual differences between the single- and double-treated arms are well below the MDE, confirming that the null result on the second upvote reflects insufficient power to detect small differences rather than evidence of exact equality. \textit{Note:} Sample sizes: Single-treated $= 3{,}287$; Double-treated $= 3{,}432$. MDE expressed in percentage points using pooled standard deviations within the two treated arms. Cohen's $d$ for the MDE at 80\% power is 0.068.}
\label{tab:s8}
\end{table}

\begin{table}[H]
\centering
\begin{tabular}{l rr rrrr}
\toprule
Outcome & Coef & SE & Raw $p$ & Bonf.\ $p$ & BH $p$ & Holm $p$ \\
\midrule
Asked another question & 0.017 & 0.007 & 0.011 & 0.034 & 0.011 & 0.011 \\
Answered another question & 0.021 & 0.006 & $<$0.001 & $<$0.001 & $<$0.001 & $<$0.001 \\
Any new post & 0.027 & 0.007 & $<$0.001 & $<$0.001 & $<$0.001 & $<$0.001 \\
\bottomrule
\end{tabular}

\vspace{4pt}
\caption{Multiplicity Adjustments for Primary Outcomes (4 Weeks). Raw $p$-values for the three pre-registered primary outcomes are adjusted using Bonferroni, Benjamini--Hochberg (BH), and Holm step-down corrections. All three outcomes survive all corrections at $\alpha = 0.05$. \textit{Note:} Panel A: All three primary outcomes remain significant at $\alpha = 0.05$ under all correction methods. Panel B: The single-treatment coefficient survives Bonferroni correction for all three outcomes. The double-treatment additional coefficient is insignificant before and after adjustment, consistent with the null effect of a second upvote.}
\label{tab:s9}
\end{table}

\newpage
\subsection*{Mechanism analyses}

Tables~\ref{tab:s10}--\ref{tab:s18} investigate how much of the treatment effect flows through the algorithmically mediated channel (upvote raises visibility, which increases answer receipt, which changes behavior) versus the direct signal. These analyses use subgroup comparisons, observational benchmarking, mediation decomposition, and heterogeneity tests to bound the contributions of each channel.

\begin{table}[H]
\centering
\small
\begin{tabular}{l cc c ccc}
\toprule
 & \multicolumn{2}{c}{Subgroup ITT} & & \multicolumn{3}{c}{Interaction Model} \\
\cmidrule(lr){2-3} \cmidrule(lr){5-7}
 & Not answered & Already answered & & Treated & Already Ans. & Interaction \\
Outcome & ($N = 18{,}564$) & ($N = 4{,}292$) & & & & \\
\midrule
New Question & 0.027$^{***}$ & $-$0.005 & & 0.031$^{***}$ & 0.066$^{***}$ & $-$0.030 \\
 & (0.007) & (0.017) & & (0.007) & (0.008) & (0.019) \\
New Answer & 0.023$^{***}$ & 0.008 & & 0.028$^{***}$ & $-$0.022$^{***}$ & $-$0.019 \\
 & (0.006) & (0.014) & & (0.006) & (0.007) & (0.015) \\
New Post & 0.039$^{***}$ & $-$0.011 & & 0.047$^{***}$ & 0.045$^{***}$ & $-$0.051$^{***}$ \\
 & (0.008) & (0.018) & & (0.008) & (0.008) & (0.019) \\
\midrule
First stage & 0.079$^{***}$ & 0.045$^{**}$ & & & & \\
 & (0.008) & (0.016) & & & & \\
\addlinespace
Equivalence (TOST $p$) & & Q: 0.005 & & & & \\
 & & A: 0.001 & & & & \\
 & & Post: 0.016 & & & & \\
\bottomrule
\end{tabular}

\vspace{4pt}
\caption{Pre-Treatment Already-Answered Split. The sample is split by whether the focal question had received at least one answer at baseline ($t_0$). The treatment effect is estimated separately within each subgroup using the pooled treatment indicator. The interaction model estimates the differential effect in the already-answered subgroup. Equivalence tests (TOST, $\delta = 0.05$) confirm that effects in the already-answered subgroup are bounded within $\pm 5$ percentage points. All models include controls for log prior questions, log prior answers, and log baseline views. Robust standard errors in parentheses. \textit{Note:} ``Not answered'' restricts to questions with zero answers at $t_0$ ($N = 18{,}564$). ``Already answered'' restricts to questions with $\geq 1$ answer at $t_0$ ($N = 4{,}292$). The first stage is the treatment effect on receiving a new answer between $t_0$ and $t_4$. Equivalence tests confirm that effects in the already-answered subgroup are within $\pm 5$ percentage points of zero at $p < 0.05$. The already-answered variable is mechanically imbalanced across treatment arms (20.7\% control, 14.7\% single, 13.8\% double) due to differential attrition of unanswered questions.}
\label{tab:s10}
\end{table}

\begin{table}[H]
\centering
\begin{tabular}{l rrr rrr rrr}
\toprule
 & & & & \multicolumn{2}{c}{New Question} & \multicolumn{2}{c}{New Answer} & \multicolumn{2}{c}{New Post} \\
\cmidrule(lr){5-6} \cmidrule(lr){7-8} \cmidrule(lr){9-10}
Window & $N$ & Control & Treated & Coef & $p$ & Coef & $p$ & Coef & $p$ \\
\midrule
4 weeks & 10,201 & 7,350 & 2,851 & 0.017$^{+}$ & 0.059 & 0.010 & 0.110 & 0.021$^{*}$ & 0.034 \\
8 weeks & 8,347 & 5,728 & 2,619 & 0.008 & 0.429 & 0.008 & 0.273 & 0.007 & 0.519 \\
12 weeks & 7,978 & 5,476 & 2,502 & $-$0.002 & 0.828 & 0.006 & 0.488 & $-$0.001 & 0.966 \\
\bottomrule
\end{tabular}

\vspace{4pt}
\caption{Never-Answered Test. Treatment effects among users whose questions received no answer at baseline and still had no answer at the endpoint. This subsample excludes the amplification channel (by conditioning on no answer receipt), so any positive treatment effect provides conservative evidence for a direct channel. The caveat is that conditioning on a post-treatment variable introduces negative selection among treated users (their questions were boosted but still received no answer), biasing treatment effects downward. \textit{Note:} $^{+}p < 0.1$, $^{*}p < 0.05$. At 4 weeks, the treatment effect on any post is significant ($p = 0.034$) and the effect on question-asking is marginally significant ($p = 0.059$), providing conservative evidence for a direct channel beyond amplification. Effects decay at 8 and 12 weeks, which is expected given the negative selection in this subsample.}
\label{tab:s11}
\end{table}

\begin{table}[H]
\centering
\small
\begin{tabular}{l rrrr rrrr rrrr}
\toprule
 & \multicolumn{4}{c}{New Question} & \multicolumn{4}{c}{New Answer} & \multicolumn{4}{c}{New Post} \\
\cmidrule(lr){2-5} \cmidrule(lr){6-9} \cmidrule(lr){10-13}
Specification & Coef & SE & & & Coef & SE & & & Coef & SE & & \\
\midrule
Raw & 0.088 & 0.008 & & & 0.205 & 0.007 & & & 0.226 & 0.008 & & \\
+ User history & 0.071 & 0.008 & & & 0.187 & 0.007 & & & 0.201 & 0.008 & & \\
+ Question quality & 0.065 & 0.008 & & & 0.185 & 0.007 & & & 0.195 & 0.008 & & \\
+ All controls & 0.065 & 0.008 & & & 0.184 & 0.007 & & & 0.195 & 0.008 & & \\
\bottomrule
\end{tabular}

\vspace{4pt}
\caption{Observational Benchmarking Among Controls. Association between receiving an answer and future activity among control users whose questions had no answer at baseline ($N = 12{,}802$). Coefficients on ``receives answer'' are reported as controls are progressively added. The association shrinks by approximately 25\% for question-asking when controls are added, indicating positive selection. The association for answering is more stable. \textit{Note:} ``Raw'' is the bivariate association. ``+ User history'' adds log prior questions and log prior answers. ``+ Question quality'' further adds log baseline views, question length, and code snippet indicator. ``+ All controls'' further adds number of comments, first-time poster indicator, and account age. All coefficients are significant at $p < 0.001$. The question-asking association drops from 0.088 to 0.065 (26\% reduction), indicating selection. The answering association is more stable (0.205 to 0.184, 10\% reduction).}
\label{tab:s12}
\end{table}

\begin{table}[H]
\centering
\small
\begin{tabular}{l cccc cccc cccc}
\toprule
 & \multicolumn{4}{c}{New Question} & \multicolumn{4}{c}{New Answer} & \multicolumn{4}{c}{New Post} \\
\cmidrule(lr){2-5} \cmidrule(lr){6-9} \cmidrule(lr){10-13}
 & \multicolumn{2}{c}{Low Views} & \multicolumn{2}{c}{High Views} & \multicolumn{2}{c}{Low Views} & \multicolumn{2}{c}{High Views} & \multicolumn{2}{c}{Low Views} & \multicolumn{2}{c}{High Views} \\
\midrule
No answer & 0.258 & & 0.235 & & 0.091 & & 0.093 & & 0.298 & & 0.275 & \\
Got answer & 0.315 & & 0.304 & & 0.272 & & 0.281 & & 0.478 & & 0.476 & \\
\addlinespace
\midrule
\multicolumn{13}{l}{\textit{Regression coefficients (with baseline controls):}} \\
\midrule
Receives answer & 0.037$^{***}$ & & & & 0.170$^{***}$ & & & & 0.155$^{***}$ & & & \\
 & (0.012) & & & & (0.010) & & & & (0.012) & & & \\
High views & 0.006 & & & & $-$0.003 & & & & 0.004 & & & \\
 & (0.009) & & & & (0.006) & & & & (0.010) & & & \\
Answer $\times$ Views & 0.010 & & & & 0.004 & & & & 0.018 & & & \\
 & (0.015) & & & & (0.013) & & & & (0.016) & & & \\
\bottomrule
\end{tabular}

\vspace{4pt}
\caption{Views vs.\ Answers Decomposition Among Controls. Mean outcomes in a 2$\times$2 classification of control users by whether their question received an answer and whether the question's view increase was above the median. Regression controls for answer receipt, high views, their interaction, and baseline covariates. Receiving an answer is the dominant predictor; views alone do not independently predict subsequent user activity. \textit{Note:} Among never-answered users ($N = 10{,}201$), the treatment significantly increases views (coefficient $= 1.57$, $p < 0.001$), but views alone do not predict subsequent participation in the regression above. The dominant predictor of future activity is whether the question received an answer, not how many views it received.}
\label{tab:s13}
\end{table}

\begin{table}[H]
\centering
\begin{tabular}{l rrrrr}
\toprule
Outcome & Total Effect & ACME & ADE & Prop.\ Mediated & $\rho_{\text{break}}$ \\
\midrule
New Question & 0.032 & 0.006 & 0.026 & 19.5\% & 0.078 \\
New Answer & 0.028 & 0.017 & 0.011 & 59.7\% & 0.244 \\
New Post & 0.047 & 0.018 & 0.029 & 38.1\% & 0.206 \\
\bottomrule
\end{tabular}

\vspace{4pt}
\caption{Causal Mediation Analysis (Imai et al., 2010). Parametric decomposition of the treatment effect through answer receipt as mediator, estimated among users whose questions had no answer at baseline ($N = 18{,}564$). ACME = average causal mediation effect; ADE = average direct effect. Sequential ignorability is almost certainly violated; these estimates are reported as a parametric benchmark, not as causal mediation. \textit{Note:} The mediator model is: receives\_answer $\sim$ treated + controls (OLS). The outcome model is: $Y \sim$ treated + receives\_answer + controls (OLS). Controls include log prior questions, log prior answers, and log baseline views. $\rho_{\text{break}}$ is the approximate correlation between mediator and outcome errors that would reduce the ACME to zero. For question-asking, a small correlation ($\rho = 0.08$) suffices, indicating that this mediation estimate is fragile. For answering, $\rho = 0.24$ is required, indicating more robustness. These values should be interpreted cautiously; they assume a specific functional form for the sensitivity analysis.}
\label{tab:s14}
\end{table}

\begin{table}[H]
\centering
\small
\begin{tabular}{l r rr rrr r}
\toprule
 & & First & & \multicolumn{3}{c}{ITT} & Wald \\
\cmidrule(lr){5-7}
Tercile & $N$ & Stage & & New Q & New A & New Post & (New Q) \\
\midrule
Low & 6,473 & 0.164$^{***}$ & & 0.045$^{***}$ & 0.047$^{***}$ & 0.062$^{***}$ & 0.276 \\
Medium & 6,173 & 0.112$^{***}$ & & 0.036$^{***}$ & 0.026$^{**}$ & 0.051$^{***}$ & 0.318 \\
High & 5,918 & $-$0.014 & & 0.019 & 0.016 & 0.034$^{*}$ & --- \\
\addlinespace
\midrule
\multicolumn{8}{l}{\textit{Continuous interaction: Y $\sim$ treated + pred\_answer + treated $\times$ pred\_answer}} \\
\midrule
 & & & & Coef & $p$ & Coef & $p$ \\
Treated & & & & 0.091 & 0.001 & 0.105 & $<$0.001 \\
Pred.\ answer & & & & 0.683 & $<$0.001 & 0.774 & $<$0.001 \\
Interaction & & & & $-$0.130 & 0.051 & $-$0.172 & 0.004 \\
\bottomrule
\end{tabular}

\vspace{4pt}
\caption{Predicted Answerability Heterogeneity. A gradient-boosted model was trained on control-group data to predict P(receives answer) from pre-treatment features. The sample (questions not answered at $t_0$) is split into terciles of predicted answerability. Under pure amplification, the Wald ratio (ITT/first stage) should be constant across terciles. \textit{Note:} The GBM uses 200 trees with max depth 3, trained on 12,802 control observations. Features: log prior questions, log prior answers, log baseline views, question length, code snippet indicator, number of comments, first-time poster indicator, account age. Tercile cutpoints: 0.380, 0.462. In the high tercile, the first stage is near zero ($-$0.014), yet the treatment effect on any post remains marginally significant. The continuous interaction is significantly negative for answering ($p = 0.004$), consistent with the treatment effect being partly proportional to the amplification channel. For asking, the interaction is marginally significant ($p = 0.051$).}
\label{tab:s15}
\end{table}

\begin{table}[H]
\centering
\begin{tabular}{l l rrr r}
\toprule
Interaction & Outcome & Coef & SE & Raw $p$ & BH $p$ \\
\midrule
Treated $\times$ log(Q length) & New Question & 0.000 & 0.009 & 0.990 & 0.990 \\
 & New Answer & $-$0.007 & 0.008 & 0.431 & 0.990 \\
 & New Post & $-$0.005 & 0.010 & 0.622 & 0.990 \\
Treated $\times$ log(Prior Qs) & New Question & $-$0.005 & 0.005 & 0.368 & 0.990 \\
 & New Answer & 0.000 & 0.005 & 0.949 & 0.990 \\
 & New Post & $-$0.002 & 0.006 & 0.731 & 0.990 \\
\bottomrule
\end{tabular}

\vspace{4pt}
\caption{Additional Heterogeneity Analyses. Interaction between treatment and question length (log) and treatment and prior experience (log prior questions). No significant heterogeneity is detected. $p$-values are corrected using Benjamini--Hochberg across all 6 tests. \textit{Note:} Each row reports the interaction term from a regression of the outcome on treatment, the moderator, and their interaction. No interaction is significant before or after FDR correction. The treatment effect does not vary with question length or prior experience.}
\label{tab:s16}
\end{table}

\begin{table}[H]
\centering
\begin{tabular}{l rrrr}
\toprule
Outcome & Coef & SE & TOST $p$ & Conclusion \\
\midrule
New Question & $-$0.005 & 0.017 & 0.005 & Equivalent \\
New Answer & 0.008 & 0.014 & 0.001 & Equivalent \\
New Post & $-$0.011 & 0.018 & 0.016 & Equivalent \\
\bottomrule
\end{tabular}

\vspace{4pt}
\caption{Equivalence Tests for Small Subgroups. Two one-sided tests (TOST) for the already-answered subgroup ($N = 4{,}292$, of which 957 treated). The equivalence margin is $\delta = 0.05$ (5 percentage points). All outcomes pass the equivalence test, confirming that the treatment effect in the already-answered subgroup is bounded within $\pm 5$ percentage points. \textit{Note:} TOST $p < 0.05$ indicates that the true effect is within $(-\delta, +\delta)$ at the 5\% level. This does not prove the effect is exactly zero, but bounds it to be small. The equivalence margin of 5 percentage points corresponds to roughly 15--30\% of the control mean depending on the outcome.}
\label{tab:s17}
\end{table}

\begin{table}[H]
\centering
\begin{tabular}{l rrrr}
\toprule
Outcome & ITT & SE & $p$ & Wald \\
\midrule
New Question (t4--t8) & 0.014$^{*}$ & 0.006 & 0.024 & 0.346 \\
New Answer (t4--t8) & $-$0.001 & 0.004 & 0.723 & $-$0.036 \\
New Post (t4--t8) & 0.009 & 0.007 & 0.171 & 0.223 \\
\midrule
\multicolumn{5}{l}{First stage: $\hat{\pi} = 0.041^{***}$ (SE $= 0.008$)} \\
\bottomrule
\end{tabular}

\vspace{4pt}
\caption{Lagged Outcomes Robustness Check (Temporal Ordering). The main analyses measure the mediator (answer receipt) and outcomes over the same 0--4 week window. To address potential temporal ordering concerns, we use 8-week data with the mediator measured at week 4 and outcomes measured between weeks 4 and 8 only. Sample restricted to questions not answered at baseline ($N = 16{,}610$). First stage ($\hat{\pi}$) is the treatment effect on answer receipt by week 4. Wald = ITT/$\hat{\pi}$. \textit{Note:} $^{*}p < 0.05$, $^{***}p < 0.001$. The asking effect persists into weeks 4--8 ($p = 0.024$), confirming that the treatment increases asking even after the mediator (answer receipt) has been realized. The answering effect is not significant in this lagged window, indicating that the answering increase is concentrated in weeks 0--4, when users return to read answers and encounter opportunities to answer. The smaller first stage (0.041 vs.\ 0.079 in the main analysis) reflects that many questions that will eventually be answered have already been answered by week 4.}
\label{tab:s18}
\end{table}

\newpage
\subsection*{Additional robustness checks}

Tables~\ref{tab:s19}--\ref{tab:s21} report three additional robustness checks. Table~\ref{tab:s19} tests whether the experimental upvote triggers cascading organic upvotes from other users (it does not). Table~\ref{tab:s20} tests whether the treatment effect is concentrated among users who cross a reputation threshold that unlocks platform privileges (it is not). Table~\ref{tab:s21} confirms that the treatment effect on views is robust to log-transformation of the ratio.

\begin{table}[H]
\centering
\begin{tabular}{l r}
\toprule
 & Organic Vote Change \\
\midrule
Treated & $-$0.009 \\
 & (0.013) \\
Log Prior Questions & 0.028$^{***}$ \\
 & (0.007) \\
Log Prior Answers & 0.048$^{***}$ \\
 & (0.008) \\
Log Prior Views & 0.046$^{*}$ \\
 & (0.018) \\
\midrule
Control mean & 0.07 \\
$N$ & 22,856 \\
\bottomrule
\end{tabular}

\vspace{4pt}
\caption{Treatment Effect on Organic Vote Changes (4 weeks). The dependent variable is the change in the question's net vote count between baseline and four weeks, excluding the experimentally assigned upvote(s). If the treatment triggered cascading organic upvotes from other users, we would expect a positive coefficient on Treated. HC1 robust standard errors in parentheses. $N = 22{,}856$. \textit{Note:} $^{*}p < 0.05$, $^{***}p < 0.001$. Treatment has no detectable effect on organic vote changes ($p = 0.494$). The experimental upvote does not trigger additional upvotes from other users, ruling out cascading organic votes as a mediator of the treatment effect.}
\label{tab:s19}
\end{table}

\begin{table}[H]
\centering
\begin{tabular}{l ccc}
\toprule
 & New Question & New Answer & New Post \\
\midrule
Treated & 0.029$^{***}$ & 0.029$^{***}$ & 0.042$^{***}$ \\
 & (0.007) & (0.006) & (0.007) \\
Crosses 15 & 0.052$^{***}$ & 0.051$^{***}$ & 0.070$^{***}$ \\
 & (0.012) & (0.009) & (0.013) \\
Treated $\times$ Crosses 15 & $-$0.031 & $-$0.035$^{*}$ & $-$0.044 \\
 & (0.021) & (0.017) & (0.023) \\
Controls & Yes & Yes & Yes \\
\midrule
$N$ & 22,856 & 22,856 & 22,856 \\
\bottomrule
\end{tabular}

\vspace{4pt}
\caption{Heterogeneity by Reputation Threshold Crossing (4 weeks). ``Crosses 15'' is an indicator for users whose baseline reputation is below 15 and who would cross the 15-point threshold (the upvote privilege) with a 10-point gain from one experimental upvote. This subgroup comprises 10.2\% of the sample. If the treatment effect were driven by mechanically unlocking platform privileges rather than by the social signal, we would expect a positive interaction. HC1 robust standard errors in parentheses. $N = 22{,}856$. \textit{Note:} $^{*}p < 0.05$, $^{***}p < 0.001$. Controls are log prior questions, log prior answers, and log baseline views. The interaction term is negative for all three outcomes and statistically significant for answering ($p = 0.034$). Users who cross the 15-point reputation threshold as a result of the experimental upvote do not drive the treatment effect; if anything, the effect is weaker for this subgroup. This argues against a purely mechanical privilege-unlocking interpretation of the results.}
\label{tab:s20}
\end{table}

\begin{table}[H]
\centering
\begin{tabular}{l cc}
\toprule
 & M1 & M2 \\
\midrule
Treated & 0.280$^{***}$ & 0.254$^{***}$ \\
 & (0.010) & (0.013) \\
Double Treated & & 0.051$^{**}$ \\
 & & (0.016) \\
\midrule
Control mean & 1.336 & 1.336 \\
$N$ & 22,856 & 22,856 \\
\bottomrule
\end{tabular}

\vspace{4pt}
\caption{Log-Transformed Views Ratio Robustness Check (4 weeks). The dependent variable is $\log(\text{Views}_{t_4} / \text{Views}_{t_0})$, the natural logarithm of the ratio of views at four weeks to views at baseline. This addresses the concern that the raw ratio is right-skewed and sensitive to outliers. HC1 robust standard errors in parentheses. $N = 22{,}856$. \textit{Note:} $^{**}p < 0.01$, $^{***}p < 0.001$. The treatment effect on views is robust to log-transformation. M2 shows a small but significant additional effect of the second upvote on log views (0.051, $p = 0.002$), consistent with a second vote providing a marginal visibility boost even though it produces no detectable effect on user behavior.}
\label{tab:s21}
\end{table}

\clearpage

\bibliography{good-question}

\end{document}